\documentclass[prb,aps,twocolumn,showpacs]{revtex4}
\usepackage{amsfonts,amsmath,bm}
\usepackage[OT4]{fontenc}

\usepackage{graphicx}

\newcommand{\bk}{b_{\bm{k}}}

\newcommand{\bkd}{b_{\bm{k}}^{\dag}}

\newcommand{\bmk}{b_{-\bm{k}}}

\newcommand{\ckl}{a_{\bm{k},\lambda}}

\newcommand{\ckld}{a_{\bm{k},\lambda}^{\dag}}

\newcommand{\sumk}{\sum_{\bm{k}}}
\newcommand{\sumkl}{\sum_{\bm{k},\lambda}}

\newcommand{\wk}{w_{\bm{k}}}

\newcommand{\omk}{\omega_{\bm{k}}}

\newcommand{\fk}{f_{\bm{k}}}

\newcommand{\gkl}{g_{\bm{k},\lambda}}

\newcommand{\kk}{\bm{k}}

\newcommand{\rl}{\rangle\!\langle}

\DeclareMathOperator{\tr}{Tr}

\begin{document}

\title{Phonon effects on the radiative recombination of excitons in
  double quantum dots}  

\author{Pawe{\l} Karwat}
\affiliation{Institute of Physics, Wroc{\l}aw University of
Technology, 50-370 Wroc{\l}aw, Poland}
\author{Anna Sitek}
\affiliation{Institute of Physics, Wroc{\l}aw University of
Technology, 50-370 Wroc{\l}aw, Poland}
\author{Pawe{\l} Machnikowski}
 \email{Pawel.Machnikowski@pwr.wroc.pl} 
\affiliation{Institute of Physics, Wroc{\l}aw University of
Technology, 50-370 Wroc{\l}aw, Poland}

\begin{abstract}
We study theoretically the radiative recombination of excitons
in double quantum dots in the presence of carrier-phonon
coupling. We show that the phonon-induced pure dephasing effects and
transitions between the exciton states strongly modify the spontaneous
emission process and make it sensitive to temperature, which may lead
to non-monotonic temperature dependence of the time-resolved
luminescence. We show also that under specific resonance conditions
the biexcitonic interband polarization can be coherently transferred to
the excitonic one, leading to an extended life time of the total
coherent polarization, which is reflected in the nonlinear optical
spectrum of the system. We study the stability of this effect against
phonon-induced decoherence.
\end{abstract}


\maketitle

\section{Introduction} 
\label{sec:intro}

Systems composed of coupled nanostructures attract much attention as
they not only open the path to new applications but also provide new
ways of controlling the 
states and dynamics of confined carriers. For instance, systems build
of two closely 
located quantum dots provide the basis for an implementation of
excitonic qubits with extended life times \cite{rolon10}, allow one to
effect conditional quantum control of carrier states
\cite{unold05},  and have been
proposed as a source of entangled photons \cite{gywat02}. Carrier
transfer in such double quantum dot (DQD) structures can also
be used for initializing the spin state of a dopant Mn atom in one of
the dots \cite{goryca09}. 

The optical properties of the DQD structures are much more complex
than those of a single quantum dot (QD), which is manifested in the
occupation and coherence dynamics observed in optical experiments
\cite{borri03,bardot05,gerardot05}. 
This is partly due to the richer structure of carrier states
\cite{bryant93,szafran01,jaskolski06,szafran08} 
but also to the interplay of a
few other factors that govern the evolution of carriers in these
systems after an
optical excitation: collective spontaneous emission (radiative
recombination of 
excitons) \cite{sitek07a},
coupling between the dots \cite{danckwerts06}, as well as phonon-induced
transitions \cite{govorov03,richter06,rozbicki08a,grodecka10} and 
dephasing \cite{roszak06a,huneke08,muljarov05}. In our previous work, 
we have investigated the effect of the 
first two of these factors \cite{sitek07a} and their role in the linear
\cite{sitek09b} and nonlinear \cite{sitek09a} response of DQDs. We
have shown, in particular, that collective effects in the spontaneous
emission of uncoupled DQDs disappear as soon as the transition energy
mismatch between the dots forming the system exceeds the spectral
line width of the fundamental transition (on the order of
micro-electron-Volts), which is well beyond current manufacturing
possibilities. However, coupling between the dots leads to pronounced
collective effects even for technologically realistic values of the
energy mismatch on the order of milli-electron-Volts. We have also
pointed out that phonon-related effects can affect the collective
spontaneous emission in these systems
\cite{machnikowski09b,machnikowski09c}. 

In this paper, we study the effects of phonon-induced transitions and
dephasing on the kinetics of collective radiative recombination
in DQDs. The interplay of these two factors has recently been
shown to determine the recombination properties of a lateral DQD
system with direct to indirect exciton tunneling
\cite{hermannstadter10}. Here, we consider a vertically stacked system
with realistic 
values of the parameters and focus on the experimentally observable
features in the system dynamics assuming the spatially direct nature
of the excitons (no electric field applied). We study the decay of
the exciton occupation in a DQD in the presence of phonons and show
that it becomes strongly dependent on temperature due to
phonon-induced thermalization of single-exciton occupations and pure
dephasing processes. For negative coupling (characteristic of tunneling),
this leads to an increase of the exciton lifetime with temperature and
to its non-monotonic behavior in a typical situation when thermally
activated non-radiative decay processes are present at higher
temperatures. We show also that for specific values of the
system parameters, coherent transfer of interband polarization between
the biexciton and exciton transitions can take place, which is
reflected in the non-linear optical spectroscopy of these systems, and
we study the stability of this process against phonon-induced decoherence.

The paper is organized as follows. In Sec.~\ref{sec:model}, we define
the model under study. Next, in Sec.~\ref{sec:evol}, we define the
method of numerical simulations and present a preliminary discussion
of the most important phenomena that will determine the evolution,
based on analytical solutions in the limiting
cases. Sec.~\ref{sec:res} contains the results of our modeling and
their discussion. Finally, Sec.~\ref{sec:concl} concludes the paper.

\section{Model}
\label{sec:model}

We consider a self-assembled structure composed of two vertically
stacked QDs with relatively similar sizes in the absence of an 
external electric field. The lowest sector of exciton eigenstates is
then formed by spatially direct states, that is, configurations in
which the electron-hole pairs reside in the same dot
\cite{szafran05,szafran08}.
We restrict
the discussion to the lowest exciton state 
in each dot and assume that
the spins of the carriers are fixed, corresponding to bright exciton
configurations. Under these assumptions, each dot can be either empty
or occupied by one exciton and the model includes four basis exciton
states: the state without 
excitons, $|0\rangle$, the states $|1\rangle$ and $|2\rangle$ with an
exciton in the first and second dot, respectively, and the ``molecular
biexciton'' state $|B\rangle$ with excitons in both dots. 
The carriers confined in the QDs
interact with the longitudinal acoustic phonon modes via the
deformation potential coupling 
and with the photon (radiative) environment via the dipole coupling. 

The total Hamiltonian of the system is
\begin{equation*}
H=H_{\mathrm{DQD}}
+H_{\mathrm{ph}}+H_{\mathrm{rad}}
+H_{\mathrm{c-ph}}+H_{\mathrm{c-rad}}.
\end{equation*}
The first term describes exciton states in the DQD structure and has
the form
\begin{eqnarray}
H_{\mathrm{DQD}}^{(\mathrm{S})} & = &   \sum_{i=1,2}\epsilon_{i}|i\rl i|
+(\epsilon_{1}+\epsilon_{2}+E_{\mathrm{B}})|B\rl B|\nonumber \\
&&
+V(|1\rl 2|+|2\rl 1|),
\label{ham-DQD-S}
\end{eqnarray}
where $\epsilon_{1}$ and $\epsilon_{2}$ are the fundamental excitonic
transition energies in the two dots,
$V$ is the coupling between the
dots, which may originate either from the Coulomb (F{\"o}rster)
interaction or from tunnel coupling (it can be assumed real),
$E_{\mathrm{B}}$ is the biexciton shift, and the superscript ``S''
denotes the Schr\"odinger picture.
Electron and hole wave functions in the dots 1 and 2 are modeled by
identical anisotropic Gaussians 
with identical extensions $l$ in the $xy$ plane
and $l_{z}$ along the growt axis $z$ for both particles,
\begin{equation}\label{wf}
\psi_{1,2}(\bm{r})\sim\exp\left[ 
-\frac{x^{2}+y^{2}}{2l^{2}} - \frac{(z\pm D/2)^{2}}{2l_{z}^{2}}
 \right],
\end{equation}
where the upper (lower) sign refers to $\psi_{1}$ ($\psi_{2}$) and 
$D$ is the distance between the dots.

The phonon modes are described by the free phonon Hamiltonian
\begin{displaymath}
H_{\mathrm{ph}}=\sumk\hbar\omk\bkd\bk,
\end{displaymath}
where $\bk,\bkd$ are the bosonic
operators of the phonon modes and $\omk$ are the corresponding
frequencies.
We assume a linear dispersion relation for phonons.
Interaction of carriers confined in the
DQD with phonons is modeled by the independent boson Hamiltonian
\begin{equation}\label{ham-phon}
H_{\mathrm{c-ph}}= \sum_{i=1,2}
(|i\rl i|+|B\rl B|)\sumk\fk^{(i)}(\bkd+\bmk), 
\end{equation}
where $\fk^{(1,2)}$
are system-reservoir coupling constants.
For Gaussian wave functions as in Eq.~\eqref{wf}, the coupling
constants for the 
deformation potential coupling between 
confined charges and longitudinal phonon modes have the form 
$\fk^{(1,2)}=\fk e^{\pm ik_{z}D/2}$, where 
\begin{displaymath}
\fk=(\sigma_{\mathrm{e}}-\sigma_{\mathrm{h}})\sqrt{\frac{k}{2\varrho vc}}
\exp\left[
-\frac{l_{z}^{2}k_{z}^{2}+l^{2}k_{\bot}^{2}}{4}\right].
\end{displaymath}
Here $v$ is the normalization volume,  
$k_{\bot/z}$ are momentum
components in the $xy$ plane and along the $z$ axis,
$\sigma_{\mathrm{e/h}}$ are deformation potential constants for
electrons/holes, $c$ is the speed of longitudinal sound,
and $\varrho$ is the crystal density. We assume that off-diagonal
carrier--phonon couplings  are negligible due to small overlap of the
wave functions confined in different dots.

The third component in our modeling is the radiative reservoir (modes
of the electromagnetic field), described by the Hamiltonian
\begin{displaymath}
H_{\mathrm{rad}}=\sumkl\hbar\wk\ckld\ckl,
\end{displaymath}
where $\ckl,\ckld$ are photon creation and annihilation operators
and $\wk$ are the corresponding frequencies ($\lambda$ denotes
polarizations). 
The QDs
are separated by a distance much smaller than the resonant
wavelength so that the spatial dependence of
the EM field may be neglected (the Dicke limit). 
The Hamiltonian describing the
interaction of carriers with the 
EM modes in the dipole and rotating wave approximations is 
\begin{equation}\label{hI-S}
H_{\mathrm{c-rad}}  = 
\Sigma_{-}\sumkl\gkl \ckld +\mathrm{H.c.},
\end{equation}
with 
\begin{displaymath}
\Sigma_{-}=|0\rl 1|+|2\rl B| + |0\rl 2| + |1\rl B|
\end{displaymath}
and
\begin{displaymath}
\gkl=i\bm{d}\cdot\hat{e}_{\lambda}(\kk)
\sqrt{\frac{\hbar\wk}{2\varepsilon_{0}\varepsilon_{\mathrm{r}}v}},
\end{displaymath}
where 
$\bm{d}$ is the interband dipole moment, $\varepsilon_{0}$ is the
vacuum permittivity, $\varepsilon_{\mathrm{r}}$ is the dielectric
constant of the semiconductor and
$\hat{e}_{\lambda}(\kk)$ is the unit polarization vector
of the photon mode with the wave vector $\kk$ and polarization
$\lambda$.  
For wide-gap semiconductors with the band gap on the order of an
electron-Volt, zero-temperature 
approximation may be used for the radiation reservoir at any
reasonable temperature. 

In our numerical simulations (to be presented in Sec.~\ref{sec:res}),
we take the parameters corresponding to a self-assembled 
InAs/GaAs system: $\sigma_{\mathrm{e}}-\sigma_{\mathrm{h}}=9$~eV,
$\rho=5350$~kg/m$^{3}$, $c=5150$~m/s, 
the wave function parameters
$l=4.5$~nm, $l_{z}=1$~nm, $D=6$~nm, and the radiative recombination time (for a
single dot) $1/\Gamma_{0}=800$~ps.

\section{Evolution}
\label{sec:evol}

The evolution of the DQD density matrix in the interaction picture is
given by the equation
\begin{equation}\label{main-evol}
\dot{\rho}=
\mathcal{L}_{\mathrm{rad}}[\rho] +\mathcal{L}_{\mathrm{ph}}[\rho],
\end{equation}
where $\rho$ is the density matrix of the exciton subsystem
and the two terms on the right-hand side account for the radiative
and phonon-induced dissipation and dephasing effects. 
Eq.~\eqref{main-evol} is valid in the leading order in both the
couplings, where the two dissipation channels are additive.

The interaction of carries with the radiation reservoir leads to a
Markovian evolution which can be described by a Lindblad
generator. For a double dot, the analysis based on the
Weisskopf--Wigner procedure \cite{sitek07a} shows that the correct
general description of the collective spontaneous emission is provided
by the equation 
\begin{displaymath}
\dot{\rho}^{(\mathrm{R})}=
-\frac{i}{\hbar}\left[
H_{\mathrm{DQD}}^{(\mathrm{R})}, \rho^{(\mathrm{R})}\right]
+\mathcal{L}_{\mathrm{rad}}^{(\mathrm{R})}[\rho^{(\mathrm{R})}],
\end{displaymath}
where
\begin{displaymath}
\mathcal{L}_{\mathrm{rad}}^{(\mathrm{R})}[\rho]
=\Gamma_{0} 
\left[ \Sigma_{-}\rho\Sigma_{+}
-\frac{1}{2}\{\Sigma_{+}\Sigma_{-},\rho \}_{+}\right].
\end{displaymath}
Here the superscript ``R'' denotes the ``rotating frame'' picture
defined by the unitary transformation 
$e^{iS(t)}$, with
\begin{equation*} 
S(t)=\frac{E}{\hbar}t (|1\rl 1|+|2\rl 2|+2|B\rl B|),
\end{equation*}
where $E=(\epsilon_{1}+\epsilon_{2})/2$. 
That is,
\begin{displaymath}
\rho^{(\mathrm{R})}=e^{iS(t)}\rho^{(\mathrm{S})}e^{-iS(t)},
\end{displaymath} 
where $\rho^{(\mathrm{S})}$ denotes the density matrix in the
Schr\"odinger picture, and
\begin{eqnarray}
H_{\mathrm{DQD}}^{(\mathrm{R})} & = & 
e^{iS(t)}H_{\mathrm{DQD}}^{(\mathrm{S})}e^{-iS(t)}-\hbar\dot{S}\nonumber\\
& = & \Delta(|1\rl 1|-|2\rl 2|)
+V(|1\rl 2|+|2\rl 1|)\nonumber\\ 
&&+E_{\mathrm{B}}|B\rl B|,
\label{ham-DQD}
\end{eqnarray}
where 
$\Delta=(\epsilon_{1}-\epsilon_{2})/2$ is the parameter describing the
transition energy mismatch between the dots. 
Since $[S(t),H_{\mathrm{DQD}}]=0$ the density matrix in the interaction
picture with respect to $H_{\mathrm{DQD}}$ is given by
\begin{displaymath}
\rho=
e^{iH_{\mathrm{DQD}}^{(\mathrm{R})}t/\hbar}\rho^{(\mathrm{R})}
e^{-iH_{\mathrm{DQD}}^{(\mathrm{R})}t/\hbar}
\end{displaymath}
and the generator of the dissipative evolution becomes
\begin{equation}\label{L-phot}
\mathcal{L}_{\mathrm{rad}}[\rho]=\Gamma_{0} 
\left[ \Sigma_{-}(t)\rho\Sigma_{+}(t)
-\frac{1}{2}\{\Sigma_{+}(t)\Sigma_{-}(t),\rho \}_{+}\right],
\end{equation}
where $\Sigma_{-}(t)=\Sigma_{+}^{\dag}(t)
=e^{iH_{\mathrm{DQD}}^{(\mathrm{R})}t/\hbar}\Sigma_{-}
e^{-iH_{\mathrm{DQD}}^{(\mathrm{R})}t/\hbar}$
and 
$\Gamma_{0}=E^{3}|\bm{d}|^{2}\sqrt{\epsilon_{\mathrm{r}}}/
(3\pi\epsilon_{0}c^{3}\hbar^{4})$
is the spontaneous decay rate for a single dot. 

For the carrier-phonon interaction, non-Markovian
effects can be important in some cases. 
Moreover, there seems to be no
universal way to extract the Markov limit in various physical
situations. 
Therefore, for the description of phonon-related effects, we use the
time-convolutionless equation in the lowest order \cite{breuer02}
\begin{equation}\label{tcl}
\mathcal{L}_{\mathrm{ph}}[\rho]=
-\int_{0}^{t}d\tau\tr_{\mathrm{ph}}\left[ 
H_{\mathrm{c-ph}}(t),\left[ H_{\mathrm{c-ph}}(\tau),\rho(t)\otimes\rho_{\mathrm{ph}} 
\right]  \right],
\end{equation}
where 
\begin{displaymath}
H_{\mathrm{c-ph}}(t)
=e^{i(H_{\mathrm{DQD}}^{(\mathrm{R})}+H_{\mathrm{ph}})t/\hbar}
H_{\mathrm{c-ph}}e^{-i(H_{\mathrm{DQD}}^{(\mathrm{R})}+H_{\mathrm{ph}})t/\hbar}
\end{displaymath}
is the carrier-phonon interaction Hamiltonian in the interaction picture,
$\rho_{\mathrm{ph}}$ is the phonon density matrix at the thermal
equilibrium, and $\tr_{\mathrm{ph}}$ denotes partial trace with
respect to phonon degrees of freedom. Note that the form of
$H_{\mathrm{c-ph}}$ is the same in the interaction and rotating frame
pictures as $[H_{\mathrm{c-ph}},S(t)]=0$.

The states
$|0\rangle$ and $|B\rangle$ are eigenstates of the DQD Hamiltonian, 
therefore nontrivial evolution generated by $H_{\mathrm{DQD}}^{(\mathrm{R})}$
takes place only in the subspace spanned by $|1\rangle,|2\rangle$. 
From the point of view of the theoretical analysis, it is convenient
to introduce the parametrization 
$\Delta=\mathcal{E}\cos \theta,\quad 
V=\mathcal{E}\sin \theta$, with $\mathcal{E}>0$ and $\theta\in[-\pi,\pi)$, 
and to treat $\mathcal{E}$ and $\theta$ as
the independent parameters of the model. Then, the single-exciton
eigenstates of the excitonic Hamiltonian can be written as
\begin{eqnarray*}
|+\rangle & = &  \cos{\frac{\theta}{2}}|1\rangle + \sin{\frac{\theta}{2}}|2\rangle,\\
|-\rangle  & = & -\sin{\frac{\theta}{2}}|1\rangle + \cos{\frac{\theta}{2}}|2\rangle,
\end{eqnarray*}
and correspond to the energies $\pm\mathcal{E}$. Thus, $2\mathcal{E}$ is
the energy splitting between the single-exciton 
eigenstates of $H_{\mathrm{DQD}}^{(\mathrm{R})}$ and $\theta$ is the mixing
angle of the single-exciton states. 

The evolution equation \eqref{main-evol} can be solved numerically at
moderate computational cost \cite{machnikowski09b}. However, some
physical insight is gained if the two evolution generators given by
Eq.~\eqref{L-phot} and Eq.~\eqref{tcl} are treated
to some extent analytically.

For the spontaneous emission, we expand the
generator \eqref{L-phot} in terms of the eigenstates of
$H_{\mathrm{DQD}}^{(\mathrm{R})}$. In the present discussion we assume
that both the energy splitting parameter $\mathcal{E}$ and the
biexciton shift $E_{\mathrm{B}}$ are much larger than the inverse
recombination time. This is reasonable as the energy mismatch and
the coupling are typically on the order of milli-electron-Volts, while the
recombination time is $\tau_{\mathrm{R}}\sim 1$~ns, yielding
$\hbar/\tau_{\mathrm{R}}\sim 1$~$\mu$eV. Then, we can neglect the
oscillating terms proportional to $e^{\pm i\mathcal{E}t/\hbar}$ and 
$e^{\pm iE_{\mathrm{B}}t/\hbar}$. This leads to the emission-induced
evolution of the exciton and biexciton occupations given 
(in the interaction picture) by
\begin{subequations}
\begin{equation}
\dot{\rho}_{\pm\pm}=
\Gamma_{\pm}\left( \rho_{BB}-\rho_{\pm\pm} \right), \quad
\dot{\rho}_{BB}=-2\Gamma_{0}\rho_{BB},
\label{evol-expl-occup}
\end{equation}
where 
$\rho_{\alpha\beta}=\langle\alpha|\rho|\beta\rangle$ and  
$\Gamma_{\pm}=\Gamma_{0}(1\pm\sin\theta)$.
The optically active coherences (optical polarizations) corresponding
to the exciton and 
biexciton transitions evolve according to 
\begin{align}
\dot{\rho}_{0\pm} & = 
-\frac{1}{2}\Gamma_{\pm}\rho_{0\pm}
+\Gamma_{\pm}\rho_{\pm B}e^{i(E_{\mathrm{B}}\mp 2\mathcal{E})t/\hbar},\label{evol-expl-pol-1} \\
\dot{\rho}_{\pm B} & = 
-\left( \Gamma_{0} +\Gamma_{\pm}/2\right) \rho_{\pm B}, \label{evol-expl-pol-2}
\end{align}
while the evolution of the ``dark'' exciton-exciton and vacuum-biexciton
coherences is described by
\begin{equation}\label{evol-expl-dark}
\dot{\rho}_{0B}=
-\Gamma_{0} \rho_{0B}, \quad
\dot{\rho}_{\pm\mp}=-\Gamma_{0}\rho_{\pm\mp}.
\end{equation}
\end{subequations}

One of the two essential effects following from
Eqs.~\eqref{evol-expl-occup}-\eqref{evol-expl-dark} is the 
appearance of the two different decay rates $\Gamma_{\pm}$. As a
consequence, one of the single-exciton states is brighter (decays faster) and the
other one is darker (decays slower) than a single exciton
state in an individual dot (from now on, the terms ``dark'' and
``bright'' will always refer to the spatial state in the DQD structure; the spin
configuration is always assumed to correspond to a bright exciton). 
For $\theta=\pm\pi/2$, one recovers the perfect 
superradiance (doubled decay rate) 
and subradiance (no radiative decay) of the two states. The former is a
precursor of the true superradiance of many identical atomic systems
\cite{gross82}. Note that this limit can be reached both for identical
dots $\Delta\to 0$ and for strongly coupled ones $|V|\gg |\Delta|$, which
makes it important also in the case of typically strongly
inhomogeneous QD systems.

The second effect is related to the oscillating term in
Eq.~\eqref{evol-expl-pol-1}. In a general case, it can be expected to
have little impact on the secular evolution
 since the two frequencies $2\mathcal{E}/\hbar$ and
$E_{\mathrm{B}}/\hbar$ are typically different. However, the
biexciton shift depends on the static dipole moment in the dots, hence
on the spatial arrangement of the carriers, and can be tuned e.g. by applying
an external axial electric field. Therefore, these two frequencies can
be tuned to resonance and this term becomes
non-negligible. Physically, this means that the transition from the
biexciton state to one of the single exciton state has the same energy
as the transition from the same single exciton state to the vacuum
state. As follows from Eq.~\eqref{evol-expl-pol-1}, in such case the
exciton coherence $\rho_{0\pm}$ is driven by the biexciton coherence
$\rho_{\pm B}$, hence coherence can be transferred from the biexciton
transition to the exciton transition. In Sec.~\ref{sec:coherence}, we will
show that this transfer leads to a considerable
modification of the decay of optical coherences in the signal and
affects the nonlinear optical spectrum of the system.

For the phonon contribution, we rewrite Eq.~\eqref{ham-phon} in the
eigenbasis of the exciton states
\begin{eqnarray}
H_{\mathrm{c-ph}} & = &  \sum_{\alpha=\pm}
(|\alpha\rl \alpha|+|B\rl B|)
\sumk\fk^{(\alpha)}(\bkd+\bmk)\nonumber \\
&&+\frac{1}{2}\sin\theta \sumk F_{\kk}(
|+\rl -|+|-\rl +|),
\label{ham-phon-2}
\end{eqnarray}
where
\begin{align}
\fk^{(+)}&=\cos^{2}\frac{\theta}{2}\fk^{(1)}+
\sin^{2}\frac{\theta}{2}\fk^{(2)},\nonumber \\
\fk^{(-)}&=\sin^{2}\frac{\theta}{2}\fk^{(1)}+
\cos^{2}\frac{\theta}{2}\fk^{(2)},
\end{align}
and
\begin{displaymath}
F_{\kk}=\fk^{(2)}-\fk^{(1)}
=-2i\fk\sin\frac{k_{z}D}{2}.
\end{displaymath}
The coexistence of carrier-phonon couplings diagonal and off-diagonal
in the carrier eigenstates in this interaction Hamiltonian makes the
evolution nontrivial \cite{grodecka10}. However, the diagonal coupling
terms in Eq.~\eqref{ham-phon-2} lead to non-Markovian pure dephasing on
a picosecond time scale, while the off-diagonal terms induce
transitions, which may be described to a good approximation in the
Markov limit and are characterized by time scales of several
picoseconds. Using this time scale separation, one can approximately
treat the two effects separately for the sake of interpreting the
simulation results to be presented in Sec.~\ref{sec:res}.

The pure dephasing dynamics is due to the lattice response to the ultrafast
change of the carrier state \cite{krummheuer02,jacak03b}. As the
phonon dynamics depends on the carrier state the lattice gets
correlated to the confined carriers and effectively extracts
information on the carrier state \cite{roszak06b}. This induces
dephasing of quantum superpositions of various exciton states, which
has been studied from the point of view of the optical response
\cite{huneke08} and of the decay of entanglement between the dots
\cite{roszak06a}. For the present discussion, the essential effect is
the dephasing of superpositions between the single exciton
eigenstates. Following the approach of 
Refs.~\onlinecite{roszak06a,roszak06c}, one finds the evolution of
the corresponding coherence (neglecting the effect of the off-diagonal
terms in Eq.~\eqref{ham-phon-2})
\begin{equation}\label{puredeph}
\langle +|\rho(t)|-\rangle = \langle +|\rho(0)|-\rangle e^{-h(t)},
\end{equation}
where 
\begin{align*}
h(t) & =\\
& 4\cos^{2}\theta\sumk\left|\frac{\fk}{\omk}\right|^{2} 
\sin^{2}\frac{k_{z}D}{2} (1-\cos\omk t)(2n(\omega_{\kk})+1) \\
&\stackrel{t\gg t_{\mathrm{ph}}}{\longrightarrow}
4\cos^{2}\theta\sumk\left|\frac{\fk}{\omk}\right|^{2}
\sin^{2}\frac{k_{z}D}{2}(2n(\omega_{\kk})+1)
\equiv h_{\infty},
\end{align*}
where  $n(\omega)$ are the phonon occupation numbers
and $t_{\mathrm{ph}}$ is the characteristic time of the
phonon-induced initial dephasing which is of the order of a few
picoseconds \cite{roszak06a}. The factor $\cos^{2}\theta$ reflects the
fact that for maximally delocalized eigenstates ($\theta=\pi/2$) the
charge distribution is the same for both of them and they cannot be
distinguished by the phonon response. 

\begin{figure}[tb]
\begin{center}
\includegraphics[width=60mm]{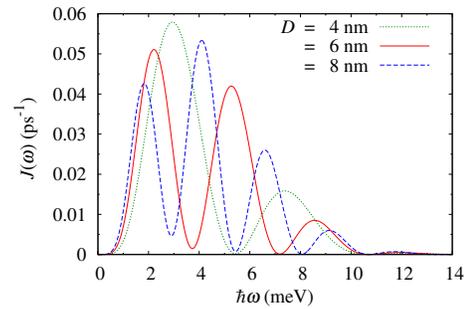}
\end{center}
\caption{\label{fig:spdens}(Color online) The phonon spectral density
  for a few values of the inter-dot distance $D$.}
\end{figure}

The real transitions governed by the off-diagonal coupling terms in
Eq.~\eqref{ham-phon-2} in many cases take place on a longer time scale
(several picoseconds to tens or hundreds of picoseconds, depending on
the amplitude of the coupling) and can be to a good approximation
described in terms of the Markovian transition rates 
$\gamma_{i\to j}=2\pi \sin^{2}\theta 
J(|\omega_{ij}|)|n(\omega_{ij})+1|$, where 
$i,j=\pm$, $\omega_{ij}=(E_{j}-E_{i})/\hbar$ 
and the spectral density is defined as
\begin{displaymath}
J(\omega)=\sumk\left| \fk \right|^{2}
\sin^{2}\frac{k_{z}D}{2}\delta(\omega-\omk).
\end{displaymath} 
The spectral density for a few values of the inter-dot spacing $D$ is
plotted in Fig.~\ref{fig:spdens}. Apart from the usual cutoff at 
frequencies larger than $\overline{l}/c$ due to the restriction on
the momentum non-conservation $\Delta(\hbar k)\lesssim
\hbar/\overline{l}$, where $\overline{l}$ is the 
average dot size, it is characterized by the oscillations with the
period $\Delta\omega=2\pi c/D$ which result from the double dot
structure of the system. These oscillations are most pronounced at
higher frequencies, where the emission along the $z$ axis (the
strongest confinement direction) dominates, which makes the factor 
$\sin^{2}(k_{z}D/2)$ most efficient.

The above remarks will provide us with useful guidelines for the
interpretation of the simulation results presented in the next section.

\section{Results}
\label{sec:res}

In this section we present and discuss the results of numerical
simulations of the spontaneous emission form double dot system under
the influence of phonons. First, in Sec.~\ref{sec:decay} we study the
decay of DQD occupations. Then, in Sec.~\ref{sec:coherence} we
describe the effect of coherent transfer of interband dipole moment
and study its stability against phonon-induced perturbations.
In all the simulations, we assume $\theta\le 0$, hence $V\le 0$ and
the lower single exciton state $|-\rangle$ is brighter than the higher
state $|+\rangle$ ($\Gamma_{-}\ge \Gamma_{+}$). 

\subsection{Occupation decay}
\label{sec:decay}

\begin{figure}[tb]
\begin{center}
\includegraphics[width=85mm]{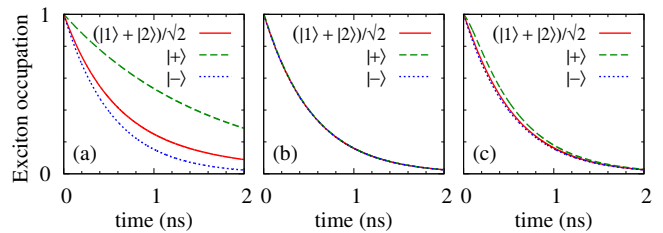}
\end{center}
\caption{\label{fig:decay}(Color online) The decay of the exciton
  occupation for three different initial states as shown for
  $D=6$~nm, $\theta=-\pi/6$, $T=4$~K.  
(a) without carrier-phonon interaction; 
(b) with phonons, for $2\mathcal{E}=1.77$~meV;
(c) with phonons, for $2\mathcal{E}=3.55$~meV.}
\end{figure}

As discussed in Sec.~\ref{sec:evol}, the coupling between the dots
leads to the formation of two 
delocalized eigenstates with different decay rates. For $V<0$, the
state $|+\rangle$ is darker (decays slower) and the state $|-\rangle$
is brighter (decays faster). This can be seen in
Fig.~\ref{fig:decay}(a) (green dashed and blue dotted lines), where we
show the evolution of the exciton 
occupation in the DQD without carrier-phonon coupling (in this case,
the evolution depends only on the mixing angle $\theta$ but not on
$\mathcal{E}$ or the system geometry). For a general
initial state, both components are present in a superposition and the
decay is not simply exponential, with an average life time in between
the two limiting cases. An example of such a situation is shown by the
red solid line in Fig.~\ref{fig:decay}(a) for the particularly
interesting initial state $(|1\rangle+|2\rangle)/\sqrt{2}$, which
corresponds to an optical excitation by a pulse that resolves
the two dots neither spatially nor spectrally.

This kinetics of recombination changes, however, when the
carrier-phonon interaction is taken into account, as shown in
Fig.~\ref{fig:decay}(b,c). Phonon-induced relaxation redistributes the
occupations on a time scale much shorter than the exciton life time,
which suppresses the dependence on the initial state. For strong
phonon coupling as in Fig.~\ref{fig:decay}(b) (corresponding to the
first maximum of the spectral density, see Fig.~\ref{fig:spdens}) the
resulting radiative decay is almost insensitive to the initial
state. The situation is still very similar in the case of energy splitting
$2\mathcal{E}$ corresponding to the first minimum of the spectral
density, which yields a much weaker
effective coupling, as shown in Fig.~\ref{fig:decay}(c). This shows
that the phonon-induced redistribution of occupations dominates the
radiative decay over a rather wide range of energy
parameters. However, for the values of $2\mathcal{E}$ around the
further minima as well as beyond the effective cut-off at about 10~meV
the phonon-related processes become inefficient and the decay becomes
similar to that observed in the absence of carrier-phonon coupling.

\begin{figure}[tb]
\begin{center}
\includegraphics[width=85mm]{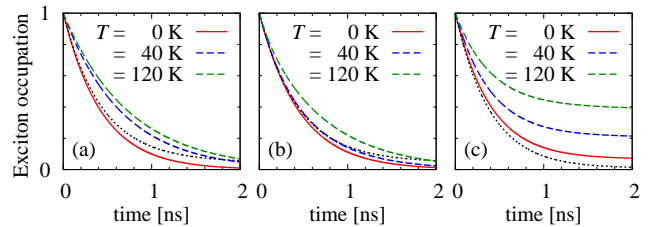}
\end{center}
\caption{\label{fig:te}(Color online) The decay of the exciton
  occupation for the initial state $(|1\rangle+|2\rangle)/\sqrt{2}$ for
  $D=6$~nm.  (a) dominated by relaxation ($\theta=-\pi/3$) with strong
  carrier-phonon coupling ($2\mathcal{E}=1.77$~meV); (b)
  as previously but with much weaker phonon effects
  ($2\mathcal{E}=7.1$~meV); (c) dominated by pure dephasing ($\theta=0$,
  $2\mathcal{E}=0$). 
}
\end{figure}

The important role of phonons in the radiative recombination leads to
a strong temperature dependence of the emission process. This effect
is shown in Fig.~\ref{fig:te}, where the exciton occupation decay for the
initial optically generated state $(|1\rangle+|2\rangle)/\sqrt{2}$ at
a few temperatures is plotted for various system parameters. 
In Fig.~\ref{fig:te}(a) we choose $\theta=-\pi/3$ and the value of the
energy splitting 
$2\mathcal{E}$ close to the first maximum of the spectral
density [Fig.~\ref{fig:spdens}], which leads to efficient
phonon-induced relaxation. The effect of the resulting redistribution of
occupations is clearly seen in the dynamics of recombination: with
increasing temperature, the occupation of the higher-energy dark state
increases at the thermal quasi-equilibrium and the resulting decay
becomes slower. Such a change of the recombination rate with
temperature can be observed also for higher value of 
$2\mathcal{E}$, corresponding to the deep second minimum of the spectral
density. However, in this case the effect is weaker and appears only
at higher temperatures, as a considerable occupation of the higher
state obviously appears only at $T\sim \mathcal{E}/k_{\mathrm{B}}$.
In Fig.~\ref{fig:te}(c) we show that a similar effect of extended life time
can appear also due to pure dephasing between the states localized in
the two dots \cite{machnikowski09b}. Here, we set $\theta=0$, which
excludes occupation redistribution. However, the
occupation-conserving pure dephasing effect remains effective and
leads to destruction of the coherence between the two single exciton
states according to Eq.~\eqref{puredeph}, where, in the present case,
$|+\rangle=|1\rangle$ and $|-\rangle=|2\rangle$. Due to this dephasing, the
initially bright state is partly turned into a mixture of dark and
bright states, in which the probability for the dark state
is\cite{machnikowski09b}  $(1-e^{-h_{\infty}})/2$ and grows with
temperature. As the lifetime of the
dark state is infinite in this limiting case of parameter values the
pure dephasing process effectively leads to a slowed down decay.

The effect of thermal extension of the exciton lifetime is directly
reflected in the dynamics of photoluminescence (PL) which may be
observed in a time-resolved
experiment. We calculate the PL signal by
multiplying the occupations of the eigenstates $|+\rangle$ and
$|-\rangle$ by the corresponding decay rates $\Gamma_{\pm}$.
In order to account for thermally activated
non-radiative processes, we follow 
Ref.~\onlinecite{bardot05} and suppress the values obtained from our
simulation by the factor $\exp(-\Gamma_{\mathrm{nr}} t)$, where 
$\Gamma_{\mathrm{nr}}=
\Gamma_{\mathrm{nr}}^{(0)}\exp[-E_{\mathrm{a}}/(k_{\mathrm{B}}T)]$ 
is the rate of
non-radiative exciton decay processes, attributed in
Ref.~\onlinecite{bardot05} to exciton escape, which are here assumed to
affect the two lowest exciton states in the same way. In this model,
$E_{\mathrm{a}}$ is the activation energy for the non-radiative
process and $\Gamma_{\mathrm{nr}}^{(0)}$ is its overall strength.
The PL decay
time is then determined as the time at which the PL intensity (which
decays not necessarily exponentially) is
reduced by a factor $1/e$. 

\begin{figure}[tb]
\begin{center}
\includegraphics[width=85mm]{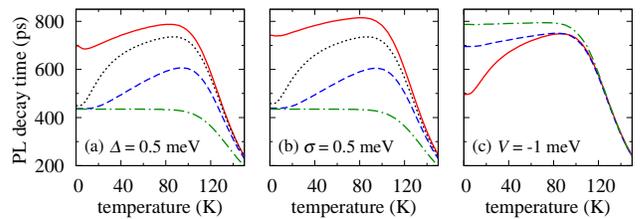}
\end{center}
\caption{\label{fig:PL}(Color online) The decay time of the DQD
  photoluminescence. (a) A single DQD with $\Delta=0.5$~meV and
  $V=-0.2$~meV (red solid line), $-1$~meV (black dotted line),
  $-3$~meV (blue dashed line), and $-8$~meV (green dash-dotted line). 
(b) An ensemble of DQDs with constant values of the coupling $V$ as in
(a), with the average energy mismatch $\overline{\Delta}=0$, and with
the standard deviation of $\Delta$ equal to
$\sigma=0.5$~meV. (c) An ensemble of DQDs for  $\overline{\Delta}=0$
and for three values the
standard deviation of $\Delta$:
$\sigma=1$~meV (red solid line), $3$~meV (blue dashed line), and
$10$~meV (green dash-dotted line).}
\end{figure}

In Fig.~\ref{fig:PL}(a) we plot the decay time of
the PL signal from a single DQD structure after initial ultrafast preparation of the state
$(|1\rangle+|2\rangle)/\sqrt{2}$ (in fact, as we discussed above, the
initial condition is not very important if the phonon-induced
thermalization is fast enough). 
We choose here a fixed value of $\Delta=0.5$~meV, 
$E_{\mathrm{a}}=95$~meV (as estimated in Ref.~\onlinecite{bardot05}) and
$\Gamma_{\mathrm{nr}}^{(0)}=8.16$~ps$^{-1}$ (half of the value in
Ref.~\onlinecite{bardot05}, corresponding to a twice shorter radiative
exciton life time assumed in our simulations). 
Varying the strength of the coupling results in a clear
qualitative change of the temperature dependence of the PL decay
time. 
The shapes of the curves obtained from the simulations  at lower
temperatures reflect the
rate of the radiative recombination, while at
higher temperatures they are
dominated by the non-radiative suppression.
In general, the thermally
increased quasi-equilibrium occupation of the higher, darker state 
leads to longer PL decay times, reduced at higher temperatures by
the non-radiative contribution, which leads to the appearance of a
maximum of the decay time. However, 
for a weak inter-dot
coupling (red solid line in Fig.~\ref{fig:PL}(a)), the energy mismatch
dominates and the two states are almost equally bright, that is, they
have similar radiative decay rates. As a result, even at $T=0$, when
only the brighter state is occupied, the decrease of the life time is
marginal and the resulting amplitude of the change in the PL decay time
is small. This changes when the coupling is stronger (black dotted and
blue dashed lines in Fig.~\ref{fig:PL}(a)) so that the ground state
has a considerably increased decay rate, which is manifested by a much
shorter PL decay time at low temperatures. In a certain range of
intermediate values of $V$ this results in a considerable growth of
the PL decay time before the non-radiative effects become important,
which leads to the formation of a pronounced maximum. However, for
higher coupling strengths 
the maximum disappears again. Now, the energy splitting between the
two states, roughly equal to $2V$, is so large that the change of equilibrium
occupations is negligible in the temperature range in which the
luminescence life time is governed by the radiative decay.

As can be seen In Fig.~\ref{fig:PL}(b), a nearly identical dependence on
temperature is obtained for an ensemble of DQDs in which the coupling
$V$ is again assumed fixed and the values of the energy mismatch $\Delta$
vary across the ensemble according to the Gaussian distribution with the
mean value $\overline{\Delta}=0$ and a rather small standard deviation
$\sigma=0.5$~meV. The fact that the ensemble of DQDs is relatively
homogeneous in terms of the energy mismatch $\Delta$ is essential. As
we show in Fig. ~\ref{fig:PL}(c), if the standard deviation $\sigma$
becomes larger the shape of the temperature dependence changes
considerably. This is due to the fact, that for $\sigma\gg |V|$ the
ensemble is dominated by DQDs with $\Delta\gg |V|$ for which the
collective effects are negligible. Hence, there is no decrease of the
PL decay time at low temperatures for such ensembles.

The temperature dependence of the PL decay time obtained from our
model is qualitatively similar to that observed in the experiment
\cite{bardot05}, where a reduction of the decay time by roughly a
factor of two at low temperatures was observed and then a maximum was
reached at about 100~K, where the decay time equal to that observed in
a reference sample with individual QDs was found. 
However, as the above
discussion shows, such a large amplitude of the variation of the PL
decay time is only possible for rather small values of $V$ (in order
to assure a small energy difference required for efficient
thermalization). Therefore, our model is unable to quantitatively
account for those experimental results obtained in a system containing
much more strongly coupled DQDs, in which thermalization of
occupations between the bright and dark states could not take place to
a sufficient extent in the relevant temperature range. In view of the
theory presented here, the origin of the strongly non-monotonic
temperature dependence of the PL decay time in such a system remains
an interesting issue. Based on our results, we can only conclude that
a thermally activated mechanism of breaking the coherent nature of the
delocalized exciton state, other than thermalization between the
two states included in our model, is needed to account for this behavior.

\subsection{Transfer of coherence}
\label{sec:coherence}

In this section we study the effect related to the resonance between
the biexciton shift $E_{\mathrm{B}}$ and the energy splitting between
the single-exciton states. As discussed in Sec.~\ref{sec:evol}, such a
resonance affects the decay of interband polarizations in the DQD
system, opening the path for a transfer of optical coherence from the
biexciton transition to the single exciton transitions, according to
Eq.~\eqref{evol-expl-pol-1}. Here, we first 
discuss this effect in the absence of phonons. Then, we describe the
phonon effects on this coherence transfer. Finally, we show that this
dynamics of interband coherences is manifested by a line narrowing in
nonlinear absorption spectra.

\begin{figure}[tb]
\begin{center}
\includegraphics[width=85mm]{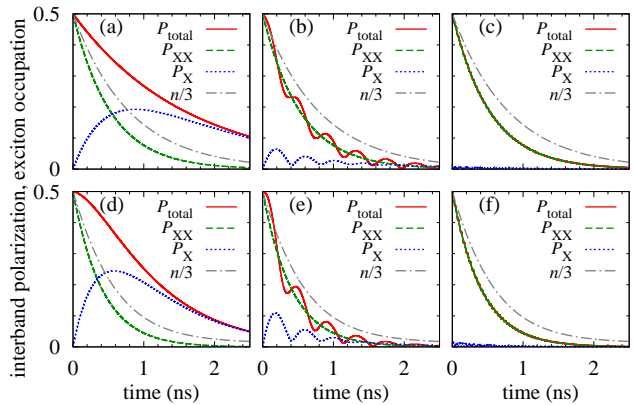}
\end{center}
\caption{\label{fig:transfer-noph}(Color online) The evolution of
  single-exciton, biexciton and total optical coherence for a DQD with
  $2\mathcal{E}=3$~meV in the absence of interaction with
  phonons. (a-c) Uncoupled dots   with 
  $E_{\mathrm{B}}=-3$~meV, $-3.01$~meV, and $-3.1$~meV,
  respectively. (d-f) Coupled dots with  $\theta=-\pi/3$ and with the
  same values of $E_{\mathrm{B}}$ as in (d-f), respectively.  In all
  the panels the red solid line shows the total 
 polarization, the green dashed line represents the biexciton
 polarization and the blue dotted line shows the single-exciton
 polarization. The evolution of the average exciton occupation (scaled
 down by a factor of 3 to fit to the polarization amplitudes) is shown
with gray dash-dotted lines for comparison.} 
\end{figure}

In Fig.~\ref{fig:transfer-noph}, we present the evolution of the
interband coherences (optical polarizations) for a single DQD in the
situation when the single-exciton transition $|-\rangle\to|0\rangle$
is energetically close to the biexciton transition
$|B\rangle\to|-\rangle$, without taking into account the
carrier-phonon interaction. We plot the evolution of the
excitonic polarization $P_{\mathrm{X}}=|\rho_{0-}|$, the biexcitonic
polarization $P_{\mathrm{X}}=|\rho_{-B}|$ and the total polarization
in the spectral vicinity of the two transitions, defined as 
$P_{\mathrm{total}}=|\rho_{0-}+\rho_{-B}|$. Figs.~\ref{fig:transfer-noph}(a-c) 
show the dynamics for uncoupled dots ($V=0$) at exact resonance,
$E_{\mathrm{B}}=-2\Delta=-3$~meV and for two slightly off-resonant
situations corresponding to  $|E_{\mathrm{B}}|-2\Delta=0.01$~meV and
0.1~meV. The initial state of the DQD is chosen to be
$(|-\rangle+|B\rangle)/\sqrt{2}$. In all the three cases, the
biexciton polarization (green dashed line) decays exactly in the same
way (with the rate $\Gamma_{0}+\Gamma_{-}/2=3\Gamma_{0}/2$ in this
case). However, at  
resonance, its decay is accompanied by the appearance of a relatively
long-living exciton polarization (blue dotted line). Hence, the decay
time of the total coherence (red solid line) is considerably
extended. This effect is present only in a narrow energy range around
the resonance. Already for a detuning of 0.01~meV, the transfer of
coherence to the single-exciton polarization is much smaller and the
decay of the total polarization is only modulated by small
oscillations. Finally, at still larger detunings from the resonance,
the coherence transfer vanishes completely and no increase of the
polarization decay time is observed. As can be seen in
Figs.~\ref{fig:transfer-noph}(d-f), the evolution is very similar for
coupled dots. Here all the time scales are slightly shorter since the
involved single-exciton state is partly superradiant. It should be
noted that the strong variation of the polarization decay time at this particular
resonance does not affect the evolution of exciton occupations, as
represented by the gray dash-dotted lines which show identical decay
independent of the detuning.

\begin{figure}[tb]
\begin{center}
\includegraphics[width=85mm]{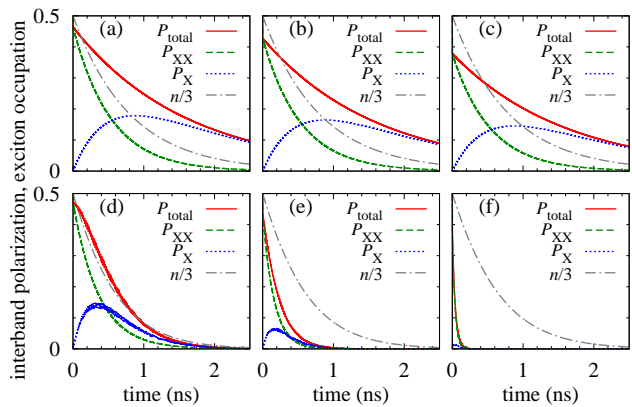}
\end{center}
\caption{\label{fig:transfer-ph}(Color online) The evolution of
  single-exciton, biexciton and total optical coherence for a DQD with
  $2\mathcal{E}=3$~meV with the carrier-phonon coupling taken into
  account. The value of $E_{\mathrm{B}}$ is taken at resonance. (a-c)
  Uncoupled dots at $T=0$, 10~K and 20~K, respectively.
 (d-f) Coupled dots with $\theta=-\pi/3$ at the same three
 temperatures. Line coding as in Fig.~\ref{fig:transfer-noph}.} 
\end{figure}

In Fig.~\ref{fig:transfer-ph}, we plot the evolution of the same
optical polarizations as in Fig.~\ref{fig:transfer-noph} but taking
into account the carrier-phonon coupling. In all the cases, the value
of $E_{\mathrm{B}}$ is set at resonance (including compensation for
phonon-induced energy shifts which are below
0.1~meV). Figs.~\ref{fig:transfer-ph}(a-c) correspond to uncoupled
dots ($V=0$) at the temperature $T=0$~K (a), 10~K (b), and 20~K
(c). In this case, the evolution of polarizations is weakly affected by
phonons due to the suppression of the phonon-induced real
transitions between the two eigenstates. The only effect is the fast
initial dephasing of the coherence \cite{krummheuer02} which increases
with temperature. In  Figs.~\ref{fig:transfer-ph}(d-f), we present
the evolution of a DQD system with coupled dots ($\theta=-\pi/3$, which
corresponds to $V\approx 1.3$~meV and $\Delta\approx 0.75$~meV). Here
we see that the lifetime of the polarization is shortened to some
extent already at $T=0$ and becomes very short at $T=20$~K. The
amplitude of the single-exciton polarization induced by the
polarization transfer also decreases with temperature. 
The effect of pure dephasing is visible also here but it affects the
evolution only over a very short period and leads to a small
almost instantaneous initial reduction of the polarization.
The strong
temperature dependence observed at longer times is due to real
phonon-induced transitions between the single exciton eigenstates
which is allowed in the case 
of coupled dots and is accompanied by a decay of coherences between
the single exciton states involved in these phonon-induced transitions
and the other two states (vacuum and biexciton). Since our initial
state involves only the lower-energy single exciton state the
phonon scattering within the single exciton subspace induces the
Markovian dephasing of the interband coherences with a rate
which critically depends on the Bose occupation factor at the phonon
frequency corresponding to the transition to the higher state, 
$n_{\mathrm{B}}(2\mathcal{E}/\hbar)$. The major phonon-induced effect
is therefore thermally activated. At higher temperatures it reduces
the lifetime of the polarizations from the usual time scales of
radiative recombination to the values typical for
phonon-related processes, that is, from almost a nanosecond to several 
picoseconds. 

\begin{figure}[tb]
\begin{center}
\includegraphics[width=85mm]{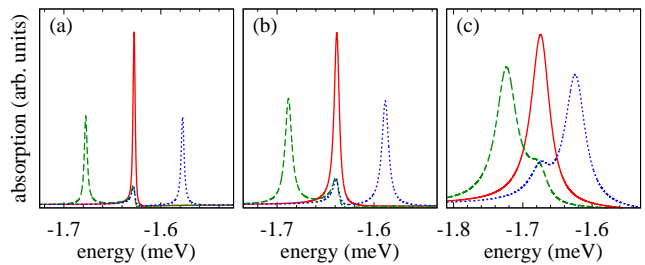}
\end{center}
\caption{\label{fig:spectra}(Color online) The pump-probe absorption
  spectra (see text) at $T=4$~K (a), $10$~K (b), and $20$~K (c). Each
  plot shows three lines corresponding to exact resonance (red solid
  lines) and to a detuning of $0.05$~meV and $-0.05$~meV from the
  resonance (blue dotted and green dashed lines, respectively).
Here $2\mathcal{E}=3$~meV, $\theta=-\pi/3$.} 
\end{figure}

While the analysis of the evolution of the individual contributions to
the polarization gives insight in the mechanisms underlying the system
kinetics, these quantities are not directly accessed in an
experiment. In order to study features related to the biexciton
coherences one usually resorts to nonlinear spectroscopy techniques
\cite{unold05,danckwerts06,kavousanaki09}. For instance, the effects
of dipole couplings between the dots have been thoroughly studied
theoretically for the 
differential absorption measured in a pump-probe
experiment\cite{danckwerts06}. Tracing the phonon effects on such spectra on the same
level of exactness would require including carrier-phonon coupling in
the full model of Ref.~\onlinecite{danckwerts06} which would go far beyond
the scope of this paper. Here, we only present a simplified model of
a particular pump-probe experiment in which the extended life times of
the optical coherences are manifested in the absorption spectra in the
most clear way. 

Let us consider a pump-probe experiment in which the spectrally
resolved lowest-order response of a single pair of coupled dots
to a probe pulse is studied after a
strong pump pulse (beyond the usual second order in the pump amplitude). 
The two pulses are spectrally broad enough to excite all the relevant
exciton and biexciton transitions in an identical way.
The delay time between the pulses is long enough for the occupations
of the two single-exciton states to equilibrate but much shorter than
the radiative life time. 
We assume that the
area of the pump pulse $\alpha_{\mathrm{pump}}$ is such that 
$\sin^{2}(\alpha_{\mathrm{pump}}/2)
=1/\{2-\tanh[\mathcal{E}/(k_{\mathrm{B}}T)]\}$. This choice strongly suppresses
the gain line corresponding to the emission from the ground exciton
state leaving the biexciton absorption line as the only feature in the
relevant spectral range. We then calculate the probe absorption
spectrum as the imaginary part of the Fourier transform of the total
polarization obtained from the simulations with the initial
state corresponding to an instantaneous excitation by the probe from the
thermalized state prepared by the pump (in the linear order in the
probe pulse area). 

The spectra obtained in this way, in the area around the transition
energy from the lowest exciton state, are shown in
Fig.~\ref{fig:spectra}. Here the 
energy is shown with respect to the average fundamental transition energy
$E=(\epsilon_{1}+\epsilon_{2})/2$. Each plot contains three spectra:
one at resonance (red solid line) and two detuned by
$\pm0.05$~meV. Each off-resonant 
spectrum contains the absorption line corresponding to the
exciton-biexciton transition and a weak feature
at the exciton-ground state transition energy. The essential effect is
the narrowing of the absorption line at the resonance,
which corresponds to the extended life time of the coherent
polarization due to the coherence transfer to single-exciton
polarization, as discussed above. This effect is quite strong at
low temperatures but becomes less pronounced when the temperature
increases and the line widths become dominated by phonon-related
effects. 

\section{Conclusion}
\label{sec:concl}

We have studied theoretically the kinetics of radiative
recombination of excitons confined in systems consisting of two
coupled quantum dots in the presence of carrier-phonon coupling. In
general terms, this corresponds to spontaneous emission from
non-identical coupled emitters undergoing an additional, possibly
non-Markovian dephasing. We 
have shown that coupling to phonons strongly affects the dynamics of
spontaneous emission by inducing transitions between the bright and
dark states of the double-dot structure as well as by pure dephasing
processes at short time scales. Both these phonon-related effects lead
to strong temperature dependence of the carrier lifetime. We have
shown that in the case of a binding ground state (negative coupling,
typical e.g. for tunneling), the carrier lifetime increases with
growing temperature. For certain values of system parameters and in
the presence of thermally activated non-radiative exciton decay
processes, this can 
lead to a non-monotonic dependence of the exciton lifetime. 

We have also studied the process of resonant coherence transfer between the
biexciton and exciton transitions and the effect of phonons on the
kinetics of this process. We have shown that this effect can lead to
considerable extension of the life time of coherent interband
polarization in the double dot system, which is stable
against phonon-induced decoherence in a relatively wide temperature
range. We have presented a simplified model of a spectrally resolved
pump-probe experiment which shows that the coherence transfer and the
resulting extension of the polarization life time manifests itself by
line narrowing in the nonlinear absorption spectrum of the system.

Our results show that carrier-phonon interaction can qualitatively
modify the optical response of
double dot systems. Phonon-related effects turn out to play a much
more important role in these systems than they do for individual
dots. Our modeling can shed some light on the existing
experimental results but also allowed us to predict new features that
can be sought for in future experiments.

\begin{acknowledgments}
This work was partly supported by the Foundation for Polish Science 
under the TEAM programme, co-financed by the European Regional
Development Fund.
\end{acknowledgments}


\end{document}